\documentclass[pre,twocolumn,amssymb,showpacs,superscriptaddress,notitlepage]{revtex4-1}

\usepackage{graphicx}
\usepackage{color}
\usepackage{epsfig}
\usepackage{latexsym}
\usepackage{bm}
\usepackage{ulem}
\usepackage{color}

\usepackage{dcolumn}
\usepackage{amsmath}    % need for subequations
\usepackage{amssymb}
\usepackage{bm} 
\usepackage{hyperref}
\usepackage{latexsym}
\usepackage{color}
\usepackage{subfigure}
\def\beq{\begin{equation}}
\def\eeq{\end{equation}}

\def\beq{\begin{equation}}                           
\def\eeq{\end{equation}}                           
\def\bea{\begin{eqnarray}}                           
\def\eea{\end{eqnarray}}        

\begin{document}

%\preprint{}

%%%%%%%%%%%%%%%%%%%%%%%%%%%%%%%%%%%%%%%%%%%%%%%%%%%
%				TITLE & ABSTRACT
\title{Polar flock in the presence of  random quenched rotators}

\author{Rakesh Das}
\email[]{rakesh.das@bose.res.in}
\affiliation{ S. N. Bose National Centre for Basic Sciences, Block JD, Sector III, Salt Lake, Kolkata 700106, India}
\author{Manoranjan Kumar}
\email[]{manoranjan.kumar@bose.res.in}
\affiliation{ S. N. Bose National Centre for Basic Sciences, Block JD, Sector III, Salt Lake, Kolkata 700106, India}
\author{Shradha Mishra}
\email[]{smishra.phy@itbhu.ac.in}
\affiliation{ Department of Physics, Indian Institute of Technology (BHU), Varanasi 221005, India}
%\date{\today}

%%%%%%%%%%%%%%%%%%%%%%%%%%%%%%%%%%%%%%%%%%%%%%%%%%%%%%%%%%%%%%%%%%%%%%%%%%%%%%%%%%%%%%%%%%%%%%%%%%%%%%%%%%%%%%%%%%%%%%%%%%%%%%%%

\begin{abstract}
We study a collection of polar self-propelled particles (SPPs) on a two-dimensional substrate in the presence of random quenched rotators.
These rotators act like obstacles which rotate the orientation of the SPPs by an angle determined by their intrinsic orientations. In the 
zero self-propulsion limit, our model reduces to the equilibrium $XY$ model with quenched disorder, while for the clean system, it is similar 
to the Vicsek model for polar flock. We note that a small amount of the quenched rotators destroys the long-range order usually noted in the 
clean SPPs. The system shows a quasi-long range order state upto some moderate density of the rotators. On further increment in the density 
of rotators, the system shows a continuous transition from the quasi-long-range order to disorder state at some critical density of rotators. 
Our linearized hydrodynamic calculation predicts anisotropic higher order fluctuation in two-point structure factors for density and velocity 
fields of the SPPs. We argue that nonlinear terms probably suppress this fluctuation such that no long-range order but only a quasi-long-range 
order prevails in the system.

{(Accepted in Phys. Rev. E (Rapid Communication))}
\end{abstract}

%\pacs {xxxxxxxxxxxxxxxxxxxxxxxxxxxx}
\maketitle

%%%%%%%%%%%%%%%%%%%%%%%%%%%%%%%%%%%%%%%%%%%%%%%%%%%%%%%%%%%%%%%%%%%%%%%%%%%%%%%%%%%%%%%%%%%%%%%%%%%%%%%%%%%%%%%%%%%%%%%%%%%%%%%%

%\section*{Introduction}
\indent Flocking of self-propelled particles (SPPs) is an ubiquitous phenomenon in nature. The size of these flocks ranges from a few 
microns to the order of a few kilometers, e.g., bacterial colony, cytoskeleton, shoal of fishes, animal herds, where the individual 
constituent shows systematic movement at the cost of its free energy. Since the seminal work by Vicsek {\it et al.} \cite{vicsek1995}, 
numerous works are done to understand the flocking phenomena of SPPs
\cite{revmarchetti2013, revtoner2005, revramaswamy2010, revvicsek2012, revcates2012}. One of the interesting features of these
kinds of out-of-equilibrium systems is the realization of true long-range order (LRO) even in two dimensions (2D) \cite{tonertu1995, tonertu1998}.
Most of the previous analytical and  numerical studies of SPPs were restricted to homogeneous or clean systems
\cite{tonertu1995, tonertu1998, chate2008, shradha2010, vicsek1995}. However, natural systems in general have some kind of 
inhomogeneity. Therefore, some of the recent studies focus on the effects of different kinds of inhomogeneities present in the systems
\cite{morin2017, chepizhko2013, marchetti2017, quint2015, sandor2017}. The study in Ref.~\cite{morin2017} shows the breakdown 
of the flocking state of artificially designed SPPs in the presence of randomly placed circular obstacles. In 
Ref.~\cite{chepizhko2013}, Chepizhko {\it et al.} model obstacles such that the SPPs avoid those obstacles. They note a surprising non-monotonicity 
in the isotropic to flocking state transition of the SPPs in the presence of the obstacles. They also report a transition from LRO to 
quasi-long-range order (QLRO) state at some nonzero but finite density of obstacles. While commenting about these studies, the 
authors of Ref.~\cite{reichhardt2017} stress upon the understanding of the flocking phenomena in the presence of different kinds 
of inhomogeneities. In the same spirit, we study the effect of rotator type obstacles on the nature of ordering in polar SPPs. Moreover, 
we propose a minimal model for SPPs in inhomogeneous medium, the results for which could easily be compared with its well-studied 
equilibrium counterpart \cite{imry1975, grinstein1976}.

In this Rapid Communication, we consider a Vicsek-like model \cite{vicsek1995} of polar SPPs in the presence of obstacles in the medium. 
The obstacles are modeled as random quenched rotators which rotate the orientation of neighboring SPPs by an angle determined by 
the intrinsic orientations of the rotators. The model can be visualized as a large moving crowd, amid which some random ``road signs" have been 
placed. Individual road sign dictates the neighboring people to take a roundabout by a certain angle from their direction of motion. 
The specific issue we address here is the correlation of this collective motion in the presence of these random road signs. 

In the limit of zero self-propulsion speed, our model reduces to the $XY$ model \cite{chaikin} with random
quenched obstacles. In the $XY$ model, any finite amount of quenched randomness is enough to destroy the orientationally ordered state 
in dimension $d \leq 4$ \cite{imry1975, grinstein1976}. Therefore in 2D, an equilibrium system with quenched obstacles 
does not have any ordered state. Analogous to this, we show that in a two-dimensional self-propelled system, quenched rotators destroy 
the LRO, usually found in the clean polar SPPs. 

In our numerical study, we note that small density of quenched rotators leads the system to a QLRO state. In this state, the absolute 
value of average normalized velocity ${\rm V}$ decreases algebraically with the system size. Also, fluctuation in the orientations 
of the SPPs increases logarithmically with system size. Moreover, below a critical density of rotators $c_{rc}$, both ${\rm V}$ 
and fluctuation in orientations of SPPs show nice scaling collapse with scaled system size. However, with further increase in density 
of rotators $c_r$, the system shows a continuous QLRO to disorder (QLRO-disorder) state transition. 
We also write hydrodynamic equations of motion for density and velocity fields of the SPPs in the presence of quenched inhomogeneities. 
A linearized study of these equations predicts an anisotropic divergence of ${\mathcal O}(1/q^4)$ in the equal-time spatially Fourier 
transformed correlations for the hydrodynamic fields for small $q$. However, neglected nonlinear terms probably suppress these 
fluctuations to make the QLRO possible in the system.

%%%%%%%%%%%%%%%%%%%%%%%%%%%%%%%%%%%%%%%%%%%%%%%%%%%%%%%%%%%%%%%%%%%%%%%%%%%%%%%%%%%%%%%%%%%%%%%%%%%%%%%%%%%%%%%%%%%%%%%%%%%%%%%%%%%%%%%%%%%%%%%%%%

%\section*{Model}\label{secmodel}
We consider a collection of $N_s$ polar SPPs distributed over a 2D square substrate. Each particle moves with a fixed 
speed ${v}_s$ along its orientation $\phi$. An individual SPP tries to reorient itself along the mean orientation of 
all the neighboring SPPs (including itself) within an interaction radius $R_s$. However, ambience noise leads to orientational 
perturbation. Moreover, there are $N_r$ immobile rotators  randomly distributed on the substrate. Each rotator possesses an 
intrinsic orientation $\varphi$, which can take any random value in the range $[-\pi,\pi]$ and remains fixed. Therefore, the rotators 
are quenched in time, and we call these random quenched rotators (RQRs). Each RQR rotates the orientations of the SPPs within 
an interaction radius $R_r$ by an angle determined by $\varphi$ and SPP-RQR interaction strength $\mu$. 
The update rules governing position ${\bm r}_i$ and orientation $\phi_i$ of the $i^{th}$ SPP are as follows:
\begin{eqnarray}
{\bm r}_i\left(t+1\right) &=& {\bm r}_i\left(t\right) + {\bm v}_i\left(t\right), \label{rupdate} \\
\phi_i\left(t+1\right) &=& \langle\phi_j\left(t\right)\rangle_{j \in R_s} + \mu\langle\varphi_j\rangle_{j \in R_r} + \Delta\psi, \label{thetaupdate}
\end{eqnarray}
where ${\bm v}_i\left(t\right) = {v}_s \left(\cos\phi_i\left(t\right),\sin\phi_i\left(t\right)\right)$ is the velocity of the 
particle $i$ at time $t$, and $\langle\phi\rangle_{R_s}$ and $\langle\varphi\rangle_{R_r}$ represent the mean orientation of all the SPPs 
and the RQRs, respectively, within the interaction radii. Fluctuation in orientation of SPPs because of ambience noise is represented 
by an additive noise term $\Delta\psi$ distributed within $\eta\left[-\pi,\pi\right]$, where noise strength $\eta \in \left[0,1\right]$. 
We call this model ``active model with quenched rotators (AMQR)," which reduces to the celebrated Vicsek model \cite{vicsek1995} 
for $\mu=0$ or in the clean system, i.e., $N_r=0$. 

%%%%%%%%%%%%%%%%%%%%%%%%%%%%%%%%%%%%%%%%%%%%%%%%%%%%%%%%%%%%%%%%%%%%%%%%%%%%%%%%%%%%%%%%%%%%%%%%%%%%%%%%%%%%%%%%%%%%%%%%%%%%%%%%%

\begin{figure}[b]
  \includegraphics[width=0.98\linewidth]{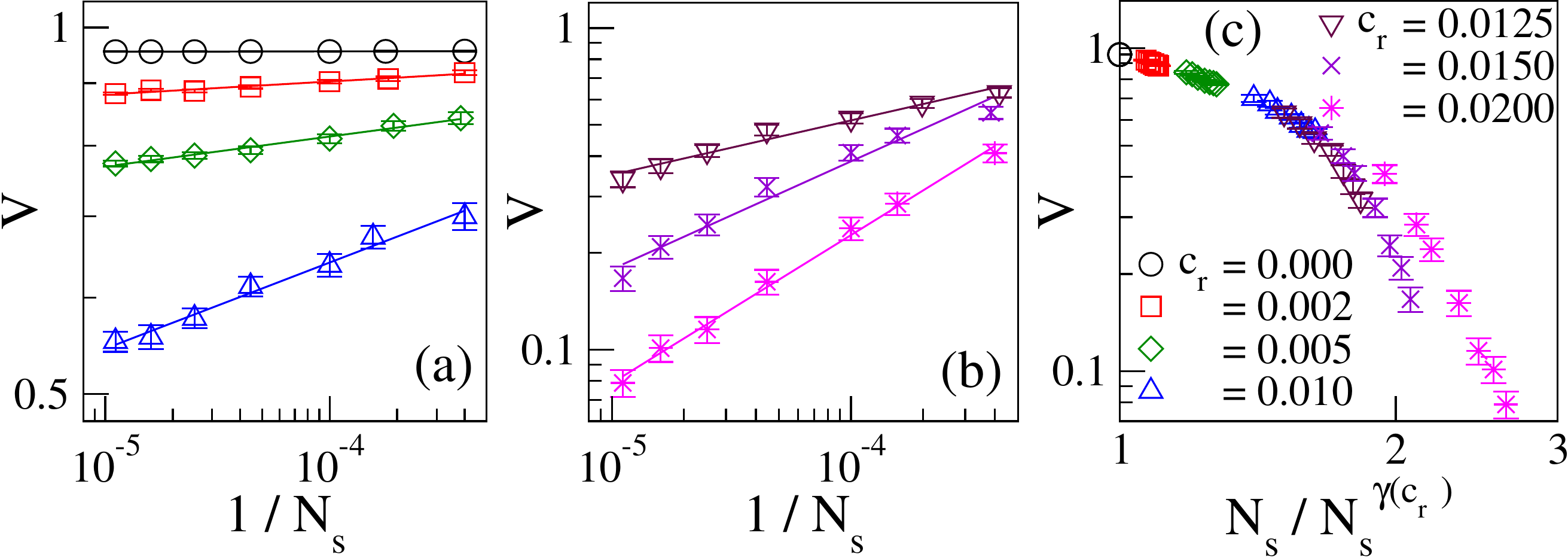}
  \caption{  ${\rm V}$ versus $1/N_s$ plots in the (a) ordered and (b) disordered state for $\eta=0.10$. 
            The error bars indicate standard error in mean. The solid lines show the respective algebraic fits.
            (c) Plot of ${\rm V}$ versus scaled system size $N_s/N_s^\gamma$ on log-log scale, where $\gamma$ is a function of $c_r$. 
            The data shows good scaling for $0<c_r\le0.0125$, but deviates for $c_r\ge0.0125$. 
          }
\label{figVNs}
\end{figure}

%%%%%%%%%%%%%%%%%%%%%%%%%%%%%%%%%%%%%%%%%%%%%%%%%%%%%%%%%%%%%%%%%%%%%%%%%%%%%%%%%%%%%%%%%%%%%%%%%%%%%%%%%%%%%%%%%%%%%%%%%%%%%%%%%

%\section*{Numerical Details and Results}\label{secnumerical}
We numerically simulate the collection of $N_s$ SPPs spread over the $L \times L$ ($L \in [50, 300]$) 2D substrate with periodic 
boundary condition. Initially the particles are chosen to have random velocity, but with constant speed ${v}_s$. The density 
of the SPPs and the RQRs are defined as $c_s=N_s/L^2$ and $c_r=N_r/L^2$, respectively. We distribute these rotators 
uniformly on the substrate, and randomly assign intrinsic orientation $\varphi \in [-\pi, \pi]$. In this system, the position and the 
velocity of all the SPPs are updated simultaneously following Eqs.~(\ref{rupdate}) and (\ref{thetaupdate}). At every time step, we use 
OpenMP Application Program Interface for a parallel updating procedure of all the SPPs. 

In this Rapid Communication, we consider $c_s=1.0$, $v_s=1.0$, and $\mu=1.0$. Moreover, we take $R_s=R_r=1$ for simplicity. In the absence of the 
rotators \cite{vicsek1995}, the system shows disorder to order transition with decreasing noise strength $\eta$. 
The ordering in the system is measured in terms of the conventional absolute value of the average normalized velocity 
\begin{equation}
{\rm V} = \langle \frac{1}{N_s {v}_s} |\sum_{i=1}^{N_s}{\bm v}_i| \rangle
\label{defvel}
\end{equation}
of the entire system \cite{vicsek1995}. Here $\langle \cdot \rangle$ indicates an average over many realizations and time in the steady state. 
${\rm V}$ varies from zero to unity for disorder to order state transition. For the reported data, we start the averaging of 
observables after $3\times10^5$ updates to assure reaching the steady state, and averaging is done for the next $5\times10^5$ updates. 
Up to $30$ realizations are used for better averaging.

%%%%%%%%%%%%%%%%%%%%%%%%%%%%%%%%%%%%%%%%%%%%%%%%%%%%%%%%%%%%%%%%%%%%%%%%%%%%%%%%%%%%%%%%%%%%%%%%%%%%%%%%%%%%%%%%%%%%%%%%%%%%%%%%%

\begin{figure}[t]
  \includegraphics[width=0.98\linewidth]{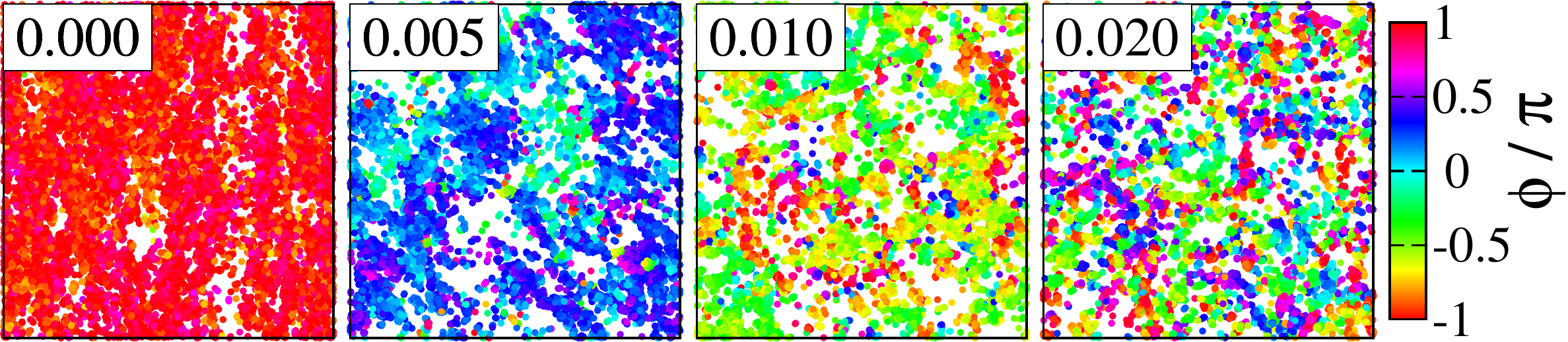}
  \caption{  Steady-state snapshots are shown for $\eta=0.10$, $L=150$ and different $c_r$ as indicated on the respective 
            panels. The color bar indicates orientation of the SPPs. The rotators with random intrinsic orientation are not 
            shown for the clarity of the figure. }
\label{figsnap}
\end{figure}

%%%%%%%%%%%%%%%%%%%%%%%%%%%%%%%%%%%%%%%%%%%%%%%%%%%%%%%%%%%%%%%%%%%%%%%%%%%%%%%%%%%%%%%%%%%%%%%%%%%%%%%%%%%%%%%%%%%%%%%%%%%%%%%%%

%{\it Quasi-long range order -} 
For a fixed $\eta$, we calculate ${\rm V}$ for different $c_r$, and study its variation with system size. 
As shown for $\eta=0.1$ in Fig.~\ref{figVNs}(a), in the clean system, ${\rm V}$ does not change with system size; consequently, the 
system possesses a nonzero ${\rm V}$ in the thermodynamic limit. Therefore, the clean system remains in the LRO state, which is a 
well-known phenomenon \cite{tonertu1998}. However, in the presence of the RQRs, ${\rm V}$ decreases algebraically with $N_s$ following 
the relation
\begin{equation}
{\rm V} = {\cal A}(c_r) N_s^{-\nu(c_r)},
\label{algebreln}
\end{equation}
as shown in Figs.~\ref{figVNs}(a) and \ref{figVNs}(b). Here both ${\cal A}$ and $\nu$ are functions of $c_r$ for a fixed $\eta$. Therefore, in the 
thermodynamic limit, ${\rm V}$ of the system with RQRs reduces to zero. We stress that for small $c_r$ the system remains in a QLRO 
state, beyond which the AMQR shows a continuous QLRO-disorder state transition, as we will see shortly. In Fig.~\ref{figsnap}, we 
show snapshots of the orientation and the local density of the SPPs for $\eta=0.1$ and different $c_r$. For $c_r=0$, all the particles 
are in highly ordered state. RQRs perturb the LRO flocking as shown for $c_r=0.005, 0.01$. For high density $c_r = 0.02$, the SPPs 
remain highly disordered. 

We further study the fluctuation in the orientation of the SPPs. The width of a normalized distribution $P(\phi)$ of orientation of 
the SPPs provides a measure of this fluctuation. It is calculated by averaging over the distributions at every time step in 
the steady state, and also over many realizations. While averaging, we set the mean orientation of all the distributions at $\phi=0$. 

%%%%%%%%%%%%%%%%%%%%%%%%%%%%%%%%%%%%%%%%%%%%%%%%%%%%%%%%%%%%%%%%%%%%%%%%%%%%%%%%%%%%%%%%%%%%%%%%%%%%%%%%%%%%%%%%%%%%%%%%%%%%%%%%%

\begin{figure}[b]
  \includegraphics[width=0.98\linewidth]{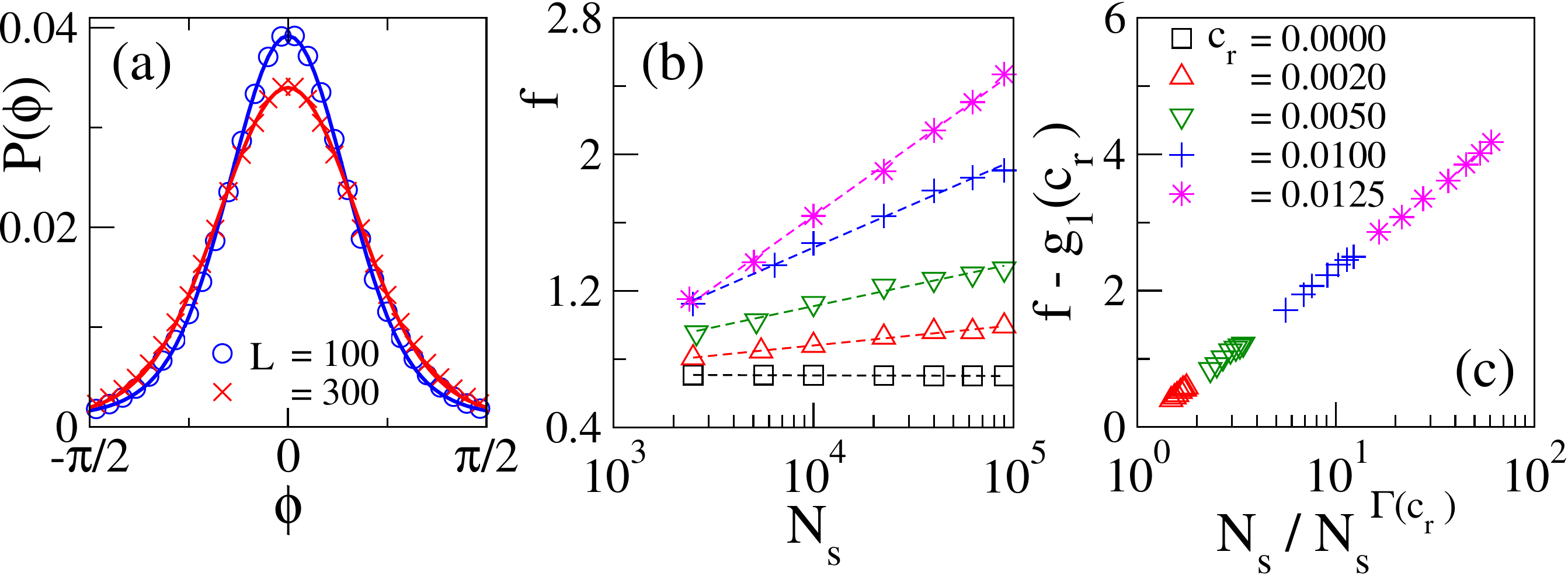}
  \caption{  (a) Distribution $P(\phi)$ of the orientation of the SPPs is shown for $\eta=0.10$ and $c_r=0.005$.
            The curves are zoomed into the range $\phi \in [-\pi/2,\pi/2]$ for better visibility. The solid lines show the respective 
            fits with Voigt profile.
            (b) Plot of the FWHM $f$ of $P({\phi})$ versus $N_s$. In the presence of quenched rotators, $f$ increases logarithmically 
            with $N_s$. The dashed lines show respective fits. 
            (c) Plot of shifted FWHM $f-g_1(c_r)$ with scaled system size $N_s/N_s^{\Gamma}$, where both $g_1$ and $\Gamma$ are 
            functions of $c_r$. The scaling holds good for $c_r \le 0.0125$. 
	  }
\label{figoriflucdir}
\end{figure}

%%%%%%%%%%%%%%%%%%%%%%%%%%%%%%%%%%%%%%%%%%%%%%%%%%%%%%%%%%%%%%%%%%%%%%%%%%%%%%%%%%%%%%%%%%%%%%%%%%%%%%%%%%%%%%%%%%%%%%%%%%%%%%%%%

We note that $P(\phi)$ widens with the increasing density of RQRs. This is quite intuitive since the degree of disorder increases 
with $c_r$. We fit these distributions with a Voigt profile, which is defined as the convolution of the Gaussian and the Lorentzian 
functions \cite{voigt}. A brief discussion of the Voigt profile and the procedure used to fit $P(\phi)$ with it are provided in 
Appendix~\ref{appVoigt}. From the respective fits, we calculate the full width at half maximum (FWHM) $f$ of the 
distributions.

We note that, in the clean system, $P(\phi)$ does not change with system size. However, for any fixed $c_r>0$, $P(\phi)$ widens 
with increasing system size, as shown in Fig.~\ref{figoriflucdir}(a) for $(\eta, c_r)=(0.10, 0.005)$ 
(also see Appendix~\ref{appVoigt}). In Fig.~\ref{figoriflucdir}(b), we show the 
variation of $f$ with system size for different $c_r$. For $c_r=0$, $f$ does not change with $N_s$. Therefore, in the clean system, 
the fluctuation in the orientation of the SPPs does not depend on the system size, and the system is in the LRO state. However, 
for $c_r>0$, FWHM of $P(\phi)$ follows the relation $f=g_1(c_r)+g_2(c_r) \ln (N_s)$, where both $g_1$ and $g_2$ are functions of 
$c_r$. 
Since $g_2 \ge 0$, $f$ increases logarithmically with $N_s$, which further confirms the QLRO in the AMQR. 

%%%%%%%%%%%%%%%%%%%%%%%%%%%%%%%%%%%%%%%%%%%%%%%%%%%%%%%%%%%%%%%%%%%%%%%%%%%%%%%%%%%%%%%%%%%%%%%%%%%%%%%%%%%%%%%%%%%%%%%%%%%%%%%%%

\begin{figure}[t]
  \includegraphics[width=0.98\linewidth]{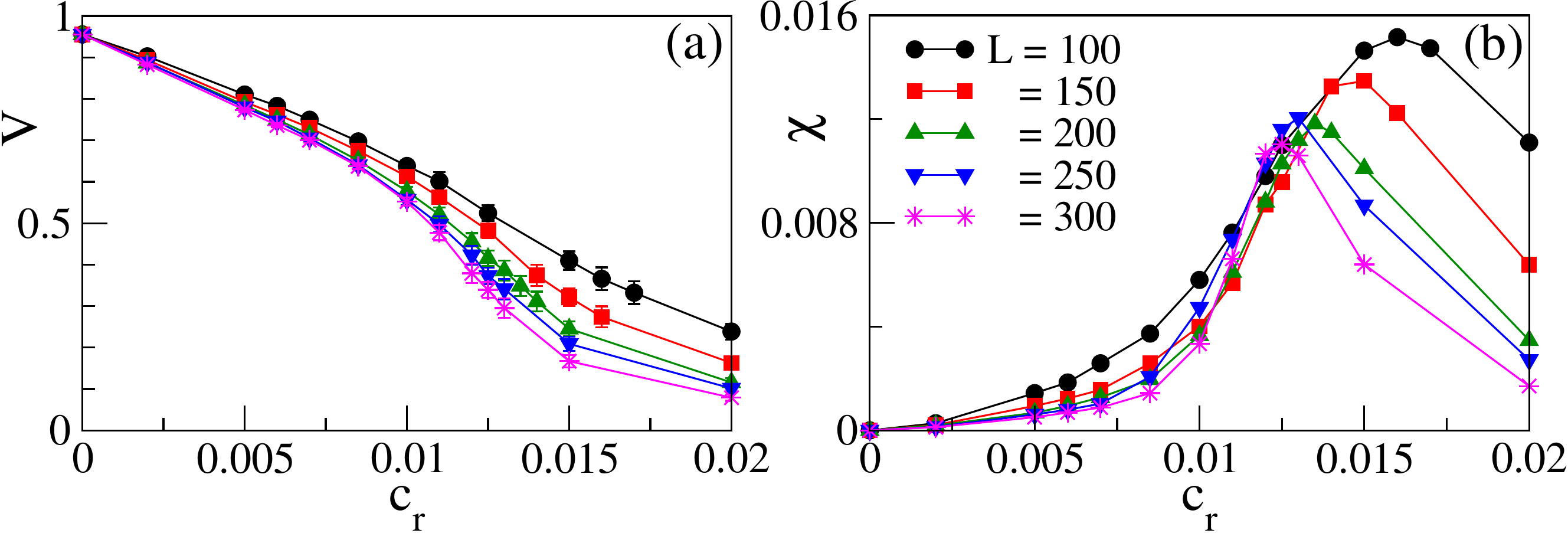}
  \caption{  (a) Variation of ${\rm V}$ with $c_r$ shown for different system sizes and $\eta=0.10$.
            (b) Variance $\chi$ of ${\rm V}$ plotted with $c_r$. The peaks in the curves indicate the critical density of the rotators
            $c_{rc}(L)$ for the QLRO-disorder transition for the respective system sizes. 
	  }
\label{figchi}
\end{figure}

%%%%%%%%%%%%%%%%%%%%%%%%%%%%%%%%%%%%%%%%%%%%%%%%%%%%%%%%%%%%%%%%%%%%%%%%%%%%%%%%%%%%%%%%%%%%%%%%%%%%%%%%%%%%%%%%%%%%%%%%%%%%%%%%%

%{\it Scaling} - 
In Fig.~\ref{figVNs}(c), we plot ${\rm V}$ versus scaled system size $N_s/N_s^{\gamma(c_r)}$ for $\eta=0.1$ and 
different $c_r$. Here $\gamma(c_r) \simeq 1-kc_r$, where $k$ is a positive constant. Moreover, $\nu = z (1-\gamma)$, where $z$ is a 
nonmonotonic function of $\eta$. We note nice scaling collapse for $c_r \le 0.0125$. This predicts that, for $c_r \le 0.0125$, the system 
can be divided into sub-systems of size $N_s^{\gamma(c_r)}$ within which the SPPs remain ordered. Since $\gamma=1$ for $c_r=0$, ${\rm V}$ 
does not depend on system size, and therefore the clean system remains in the LRO state. However, in the presence of RQRs, the 
system remains in the QLRO state. Moreover, the scaling predicts self-similarity of the system for different $c_r \le 0.0125$. As shown 
in Fig.~\ref{figoriflucdir}(c), we also find nice scaling collapse of $f-g_1(c_r)$ with scaled system size $N_s/N_s^{\Gamma(c_r)}$ for 
different $c_r \le 0.0125$, where $\Gamma = 1 - g_2$ that varies  linearly with $c_r$, for small $c_r$. 
Similar scaling holds for other $\eta$ values in the QLRO state.

\begin{figure}[b]
  \includegraphics[width=0.98\linewidth]{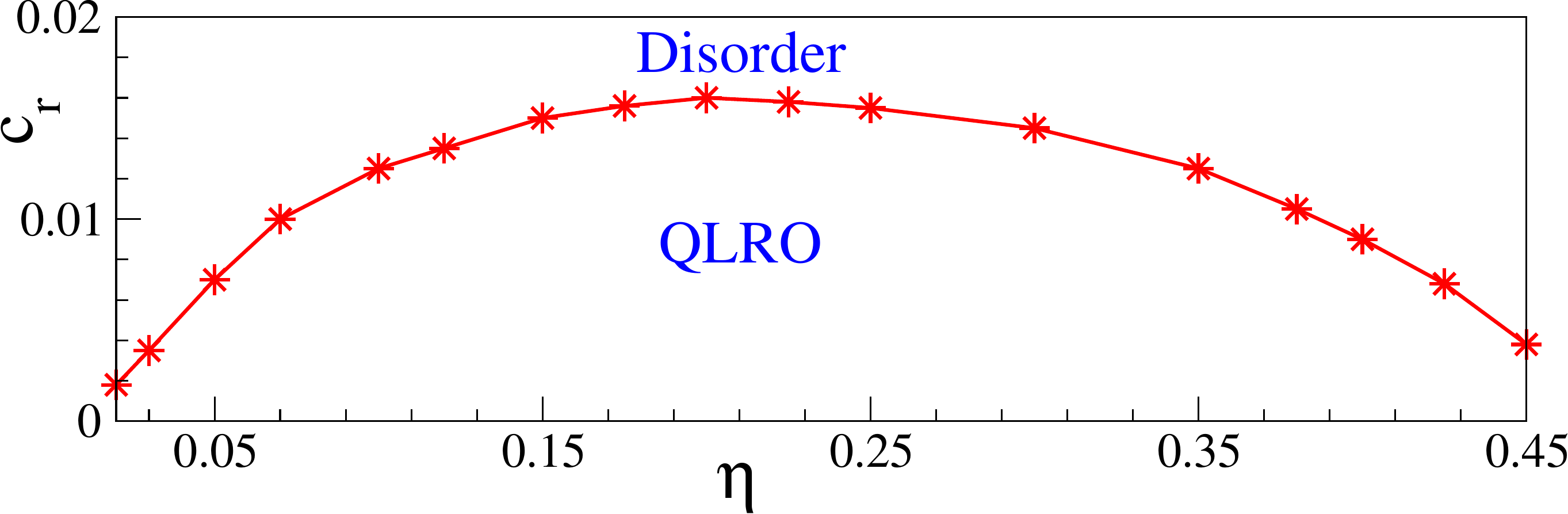}
  \caption{  Phase diagram in noise strength versus density of rotator plane. For small $c_r$, the QLRO state 
            prevails, beyond which the system continuously goes to the disorder state. }
\label{figphasediag}
\end{figure}

%{\it QLRO to disorder state transition} - 
In Fig.~\ref{figchi}(a), we show the variation of ${\rm V}$ with $c_r$ for $\eta=0.1$ and different system sizes. 
Starting from the value of ${\rm V}$ close to $1$ for small $c_r$, ${\rm V}$ shows a transition to smaller values with increasing $c_r$. 
Therefore, with increasing $c_r$, QLRO-disorder transition occurs in the system. We further calculate the variance $\chi$ of ${\rm V}$ 
for different system sizes, and plot these as a function of $c_r$ in Fig.~\ref{figchi}(b). Data shows systematic variation in $\chi$ as 
a function of $c_r$, and a peak appears at $c_r = c_{rc}(L)$ where the fluctuation in ${\rm V}$ is large. This suggests a continuous 
QLRO-disorder state transition in the AMQR. We consider $c_{rc}(L)$ as the critical density for the QLRO-disorder state transition for 
system size $L$. The position of the peak shifts from $c_r= 0.016$ to $0.0125$ as $L$ is increased from $100$ to $300$. However, we note 
that $c_{rc}(L)$ flattens on increasing $L$ for all $\eta$ values. Using the extrapolated values $c_{rc}(L \rightarrow \infty)$, we 
construct a phase diagram in the $\eta$--$c_r$ plane. We stress that in the presence of RQRs, the system remains in the QLRO below the 
phase boundary shown in Fig.~\ref{figphasediag}. 

%%%%%%%%%%%%%%%%%%%%%%%%%%%%%%%%%%%%%%%%%%%%%%%%%%%%%%%%%%%%%%%%%%%%%%%%%%%%%%%%%%%%%%%%%%%%%%%%%%%%%%%%%%%%%%%%%%%%%%%%%%%%%%%%%

%%%%%%%%%%%%%%%%%%%%%%%%%%%%%%%%%%%%%%%%%%%%%%%%%%%%%%%%%%%%%%%%%%%%%%%%%%%%%%%%%%%%%%%%%%%%%%%%%%%%%%%%%%%%%%%%%%%%%%%%%%%%%%%%%

%\section*{Hydrodynamic analysis}\label{sechydro}
%{\it Linearised hydrodynamics} -
Long-distance and long-time properties of the SPPs with quenched obstacles can also be characterized using a hydrodynamic description of 
the model. The relevant hydrodynamic variables for this model are (i) SPP density $\rho({{\bm r}, t})$ which is a globally conserved 
quantity and (ii) velocity ${\bm v}({{\bm r}, t})$ which is a broken-symmetry parameter in the ordered state. These variables can 
be obtained by suitable coarsening of corresponding discrete variables in the microscopic model 
\cite{tonertu1995, tonertu1998, aparna2008, bertin2006, bertin2009, ihle2011}. 
Following the phenomenology of the system, we write the hydrodynamic equations of motion for the density and the 
velocity fields as
\begin{eqnarray}
 \partial_t \rho &+& \nabla \cdot ({\bm v}\rho) = D_{\rho} \nabla^2 \rho, \label{eqmden} \\
 \partial_t {\bm v} &+& \lambda_1 ({\bm v} \cdot \nabla) {\bm v} + \lambda_2 (\nabla \cdot {\bm v}) {\bm v} + \lambda_3 \nabla({v}^2) \notag \\ 
                    &=& (\alpha_1 - \alpha_2 {v}^2) {\bm v} - \nabla P + D_B \nabla (\nabla \cdot {\bm v}) \notag \\
                   && \quad + D_T \nabla^2 {\bm v} + D_2 ({\bm v} \cdot \nabla)^2 {\bm v} + \frac{\rho_o}{\rho} {\bm \zeta} + {\bm f}.  \label{eqmvel}
\end{eqnarray}
${\bm f}$ represents the annealed noise term that provides a random driving force. We assume this to be a white Gaussian noise with the correlation
\begin{equation}
\langle f_i({\bm r}, t) f_j({\bm r}^\prime, t^\prime) \rangle = \Delta \delta_{ij} \delta({\bm r}-{\bm r}^\prime) \delta(t-t^\prime),  
 \label{fcorr}
\end{equation}
where $\Delta$ is a constant, and dummy indices $i, j$ denote Cartesian components. The effect of obstacles is contained in the term 
$\frac{\rho_o}{\rho} {\bm \zeta}$ in Eq.~(\ref{eqmvel}), where $\rho_o$ represents obstacle density, and ${\bm \zeta}({\bm r},t)$ 
signifies the obstacle field. We assume the correlation
\begin{equation}
\langle \zeta_i({\bm r}, t) \zeta_j({\bm r}^\prime, t^\prime) \rangle = \zeta^2 \delta_{ij} \delta({\bm r}-{\bm r}^\prime),
 \label{zetacorr}
\end{equation}
which contains no time dependence, and therefore represents a quenched noise. Equations~(\ref{eqmden})-(\ref{zetacorr}) represent the 
Toner-Tu \cite{tonertu1998} model for $\zeta =0$.

We check whether a broken-symmetry state of the SPPs in the presence of the obstacle field survives to small fluctuation in the 
hydrodynamic fields. In the hydrodynamic limit, a linearized study of Eqs.~(\ref{eqmden}) and (\ref{eqmvel}) 
gives spatially Fourier transformed equal-time correlation functions for the density 
\begin{eqnarray}
 C_{\rho\rho}({\bm q},t) = 
    \frac{1}{q^2} \left\{ \frac{\zeta^2 \rho_o^2 a_{\rho}(\theta)}{b(\theta)q^2+d(\theta)} + \Delta A_{\rho}(\theta) \right\}
 \label{Crhorho2}
\end{eqnarray}
and the velocity
\begin{eqnarray}
 C_{{vv}}({\bm q},t) = 
     \frac{1}{q^2} \left\{ \frac{\zeta^2 \rho_o^2 a_{v}(\theta)}{b(\theta)q^2+d(\theta)} + \Delta A_{v}(\theta) \right\}.
 \label{Cvv2}
\end{eqnarray}
The parameters $a_{\rho, {v}}$, $A_{\rho, {v}}$, $b$, and $d$ depend on the specific microscopic model and the angle $\theta$ between the wave 
number ${\bm q}$ and the flocking direction. A detailed calculation for Eqs.~(\ref{Crhorho2}) and (\ref{Cvv2}) is given in 
Appendix~\ref{appA}.Our result matches with the earlier prediction by Toner and Tu \cite{tonertu1998} for $\zeta = 0$, where the two 
structure factors diverge as $1/q^2$ for small $q$. 
However, the linearized theory suggests $C_{\rho\rho, vv} \sim 1/q^4$ for $\zeta \neq 0$, provided $d(\theta)=0$. In general 
for a Vicsek-like model as our AMQR, $d(\theta)$ vanishes for certain directions $\theta = \theta_c$ or $\pi - \theta_c$, where 
$\theta_c$ depends on the model parameters. We stress that although the quenched inhomogeneities increase fluctuation in the 
system as compared to the clean case, the neglected nonlinearities suppress these higher order fluctuations so that a QLRO state 
can prevail. Alhough an exact nonlinear calculation is {\it not practically feasible} for the 2D polar flock \cite{toner2018}, 
presumption of convective nonlinearities as relevant terms offers a way out \cite{revtoner2005, tonertu1998}. A nonlinear calculation 
\cite{toner2018} following this presumption renormalizes diffusivities as $1/q$ so that the term $b(\theta)q^2$ in 
Eqs.~(\ref{Crhorho2}) and (\ref{Cvv2}) approaches a finite value, and therefore, a QLRO state exists in the system. This explanation is 
consistent with the giant number fluctuation \cite{revramaswamy2010} in the AMQR. We have checked that inclusion of the RQRs increases 
the fluctuation in the system as compared to the clean case. This enhanced fluctuation destroys the usual LRO of the clean system. 
However, we note that the fluctuation decreases with further increase in $c_r$ which disagrees with Eq.~(\ref{Crhorho2}), as the 
linearized hydrodynamics prescribes an increase in the effect of quenched inhomogeneity with $\rho_o$. Therefore, the neglected 
nonlinearity indeed plays a pivotal role in stabilizing the QLRO state in the system. A detailed discussion of these phenomenologies 
is given in Appendix~\ref{appnonlinear}.

%%%%%%%%%%%%%%%%%%%%%%%%%%%%%%%%%%%%%%%%%%%%%%%%%%%%%%%%%%%%%%%%%%%%%%%%%%%%%%%%%%%%%%%%%%%%%%%%%%%%%%%%%%%%%%%%%%%%%%%%%%%%%%%%%%%%%%%%%%%%%%%%%%

%\section*{Discussion}\label{secsummary}
In summary, we have studied the effect of random quenched rotators on the flocking state of polar SPPs. These rotators are one kind of 
obstacle that rotate the orientation of the SPPs. We find that, for small density of the rotators, the usual LRO of the clean polar 
SPPs is destroyed, and a QLRO state prevails. With further increase in density of the rotators, a continuous QLRO to disorder state 
transition takes place in the system. Our linearized hydrodynamic analysis predicts an anisotropic higher order fluctuation which destroys 
the usual LRO of the clean SPPs. However, the neglected nonlinearities suppress these fluctuations yielding a QLRO in the system. In 
equilibrium systems with random quenched obstacles, an ordered state does not exist below four dimensions \cite{imry1975,grinstein1976}. 
However, as compared to the equilibrium systems, in our model for polar SPPs with quenched rotators, we find QLRO in two-dimensions. 
Our prediction of the QLRO in the polar SPPs in the presence of quenched obstacles agrees with recent observations \cite{toner2018, chepizhko2013}. 

In contrast to the LRO and the QLRO reported in Ref.~\cite{chepizhko2013}, we note QLRO only, because of the basic difference in the 
nature of obstacles. The SPP-obstacle interaction in Ref.~\cite{chepizhko2013} depends on the angle between their relative position 
vector and the orientation of the SPP. Therefore, this force is a continuous function of the orientation distribution of the SPPs. On 
the contrary, the quenched force offered by the obstacles in our model is random and discrete. However, similar to their results, 
we note the existence of an optimal noise for which the system attains the maximum ordering in the presence of quenched rotators 
(see Appendix~\ref{appoptnoise}). Our model can be applied in natural systems like a shoal of fishes moving in the sea in 
the presence of vortices. An experiment on a collection of fishes living in a shallow water pool 
\cite{jolles2017, couzin2008, krause2018, puckett2018}, in the presence of uncorrelated artificial vortices, may verify our predictions. 

%%%%%%%%%%%%%%%%%%%%%%%%%%%%%%%%%%%%%%%%%%%%%%%%%%%%%%%%%%%%%%%%%%%%%%%%%%%%%%%%%%%%%%%%%%%%%%%%%%%%%%%%%%%%%%%%%%%%%%%%%%%%%%%%%%%%%%%%%%%%%%%%%%

%\begin{acknowledgements}
S.M. acknowledges Sriram Ramaswamy for pointing out an important correction in the hydrodynamic calculation and Sanjay Puri for useful discussions. 
The authors thank John Toner for his useful comments and suggestions. S.M. also thanks S. N. Bose National Centre for Basic Sciences, Kolkata 
for providing kind hospitality, and the Department of Science and Technology, India for financial support. M.K. acknowledges financial support 
from the Department of Science and Technology, India under the Ramanujan Fellowship.
%\end{acknowledgements}

%%%%%%%%%%%%%%%%%%%%%%%%%%%%%%%%%%%%%%%%%%%%%%%%%%%%%%%%%%%%%%%%%%%%%%%%%%%%%%%%%%%%%%%%%%%%%%%%%%%%%%%%%%%%%%%%%%%%%%%%%%%%%%%%%

\appendix

\section{Voigt profile and orientation fluctuation of Self-propelled particles} \label{appVoigt}
Voigt profile is defined as 
\begin{equation}
{\cal V}(\phi; \sigma, \epsilon) = \int_{-\pi}^{\pi} \frac{\exp(-\Phi^2 / 2\sigma^2)}{\sigma\sqrt{2\pi}}
                                   \frac{\epsilon}{\pi\left[ (\phi-\Phi)^2 + \epsilon^2 \right]} d\Phi,
\label{eqvoigt}
\end{equation}
where the Gaussian and the Lorentzian contributions are signified by the parameters $\sigma$ and $\epsilon$, respectively. 
The full width at half maximum (FWHM) of the Voigt profile is approximately given by the relation \cite{voigt}
\begin{equation}
f \approx 0.5346 f_L + \sqrt{0.2166 f_L^2 + f_G^2},
\label{fwhmvoigt}
\end{equation}
where $f_L=2\epsilon$ represents the FWHM of the Lorentzian distribution, and $f_G=2\sigma\sqrt{2\ln2}$ represents the FWHM 
of the Gaussian distribution. 

As mentioned in the main text, we realise that the distribution $P(\phi)$ follows Voigt profile. So we take discrete 
Fourier transform (DFT) of $P(\phi)$ and fit the transformed distribution with the characteristic function 
${\xi}\left(n;\sigma,\epsilon\right) = \exp\left(\frac{\sigma^2 n^2}{2} - \epsilon |n|\right)$ of ${\cal V}(\phi; \sigma, \epsilon)$.
Here $n$ represents the Fourier conjugate of $\phi$. From the fits in the Fourier space, we extract the values of the parameters 
$\sigma$ and $\epsilon$, and calculate the FWHM of $P(\phi)$ using Eq.~(\ref{fwhmvoigt}). The fits shown in Fig.~3(a) of the main text
are obtained by the inverse DFT of the fitted functions ${\xi}\left(n;\sigma,\epsilon\right)$.

%%%%%%%%%%%%%%%%%%%%%%%%%%%%%%%%%%%%%%%%%%%%%%%%%%%%%%%%%%%%%%%%%%%%%%%%%%%%%%%%%%%%%%%%%%%%%%%%%%%%%%%%%%%%%%%%%%%%%%%%%%%%%%%%%

\begin{figure}[b]
  \includegraphics[width=0.9\linewidth]{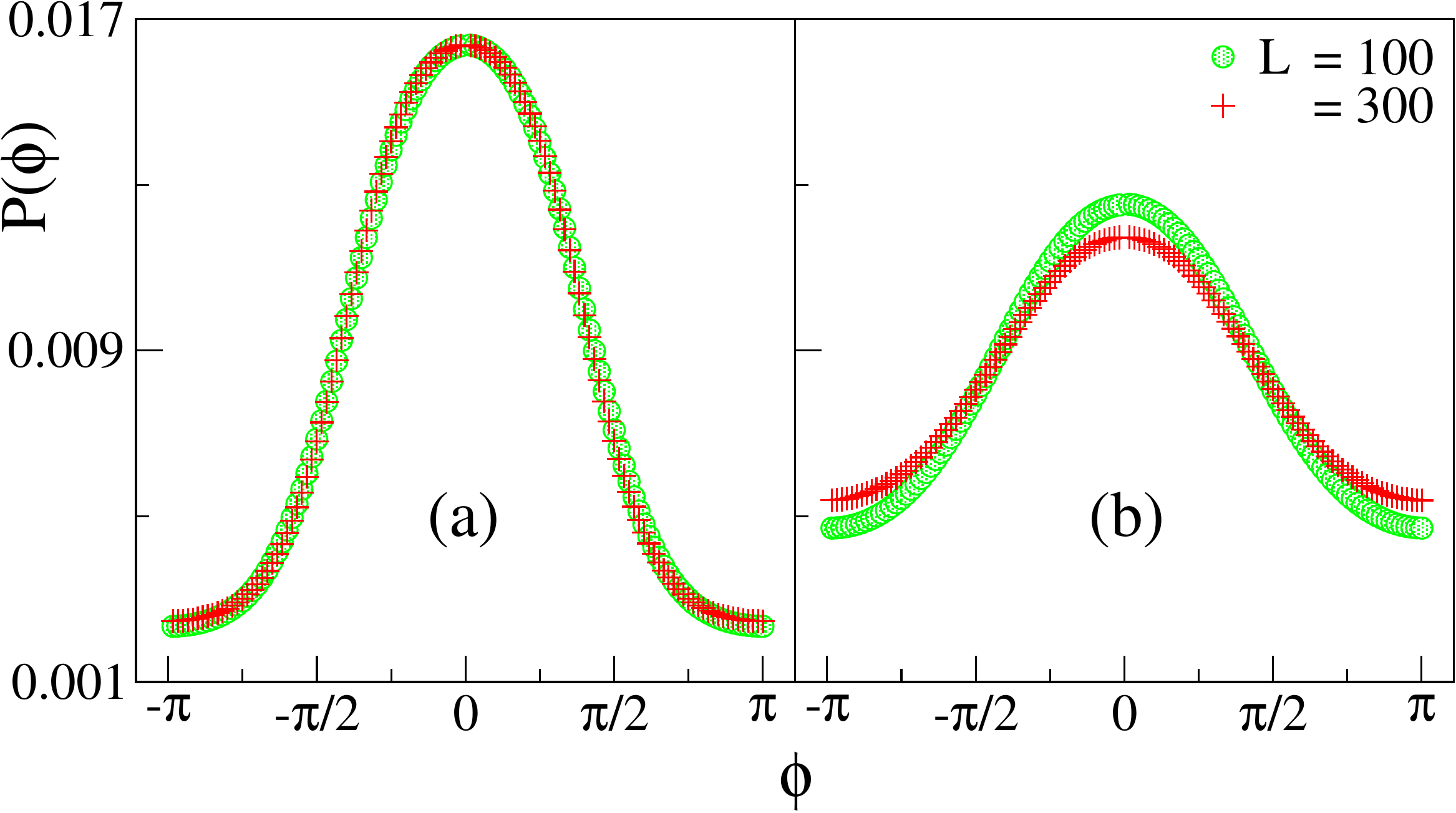}
  \caption{  Plot of orientation distribution $P(\phi)$ of the SPPs for $\eta=0.40$. In the clean system, i.e., 
            for $c_r=0$, the system remains in the banded state. As shown in (a), $P(\phi)$ does not change with system size 
            in this state. However, as shown in (b) for $c_r=0.008$, fluctuation in ${\rm V}$ increases with system size. 
	    Respective fits with the Voigt profile has not been shown for the clarity of the plots.}
\label{figbanded}
\end{figure}

%%%%%%%%%%%%%%%%%%%%%%%%%%%%%%%%%%%%%%%%%%%%%%%%%%%%%%%%%%%%%%%%%%%%%%%%%%%%%%%%%%%%%%%%%%%%%%%%%%%%%%%%%%%%%%%%%%%%%%%%%%%%%%%%%

We must stress here that, in the clean system, $P(\phi)$ is always independent of system size; no matter the system is in the 
homogeneous ordered state or in the banded state. This is evident from Fig.~\ref{figbanded}(a) where we plot $P(\phi)$ for the 
banded state ($\eta=0.40$). However, in the presence of random quenched rotators, $P(\phi)$ widens with system size, as shown 
in Fig.~\ref{figbanded}(b).

%%%%%%%%%%%%%%%%%%%%%%%%%%%%%%%%%%%%%%%%%%%%%%%%%%%%%%%%%%%%%%%%%%%%%%%%%%%%%%%%%%%%%%%%%%%%%%%%%%%%%%%%%%%%%%%%%%%%%%%%%%%%%%%%%

\section{Linearised theory of the broken symmetry state in the presence of quenched inhomogeneities} \label{appA}
Given the equations of motion (EOMs) of the hydrodynamic fields in the main text, we check whether a broken symmetry state of the SPPs in the 
presence of obstacle field survives to small fluctuations in the hydrodynamic fields. We consider a broken symmetry state 
${\bm v} = {v}_0 \hat y + \delta{\bm v}$, where the spontaneous average value of the velocity $\langle {\bm v} \rangle = {v}_0 \hat y$ 
and ${v}_0 = \sqrt {\alpha_1 / \alpha_2}$. Fluctuation in the density field is given by $\delta\rho = \rho - \bar\rho$, where $\bar\rho$
represents the mean density of SPPs. We expand spatial and temporal gradients appearing in the EOMs, and retain upto lowest-order 
terms in derivatives, since we are interested in long-time and long-distance behavior of the system. Doing so, we obtain the EOM for the 
fluctuation $\delta{v}_y$ as
\begin{equation}
 \partial_t \delta{v}_y + 2\alpha_1 \delta{v}_y = -\sigma_1 \partial_y \delta \rho + \frac{\rho_o}{\bar\rho} \zeta_y + \mbox{irrelevant terms}. 
 \label{deltdelvy}
\end{equation}
Since we are interested in hydrodynamic modes, i.e., modes for which frequency $\omega \rightarrow 0$ as wave number $q \rightarrow 0$,
we can neglect time-variation of $\delta {v}_y$ as compared to its value. Therefore, from Eq.~(\ref{deltdelvy}), we obtain the relation
\begin{equation}
 \delta {v}_y = \frac{1}{2\alpha_1} \left( -\sigma_1 \partial_y \delta\rho + \frac{\rho_o}{\bar\rho} \zeta_y \right).
 \label{delvy}
\end{equation}
Using the expression for $\delta{v}_y$ from Eq.~(\ref{delvy}), we obtain the EOMs for $\delta \rho$ and $\delta {v}_x$ as
\begin{eqnarray}
 \left( \partial_t +{v}_0 \partial_y - D_{\rho y} \partial_y^2 - D_{\rho} \partial_x^2 \right) \delta \rho + 
      \bar\rho \partial_x \delta {v}_x &=& -\frac{\rho_o}{2\alpha_1} \partial_y \zeta_y, \notag \\ \label{eqmdelrho}\\
 \sigma_1 \partial_x \delta \rho + \left( \partial_t + \gamma \partial_y - D_L \partial_x^2 - D_y \partial_y^2 \right) 
    \delta {v}_x &=& \frac{\rho_o}{\bar\rho} \zeta_x + f_x, \notag \\ \label{eqmdelvx}
\end{eqnarray}
where $D_{\rho y}=D_{\rho} + \bar\rho \sigma_1 / 2\alpha_1$, $D_L = D_B + D_T$, $D_y = D_T + D_2 {v}_0^2$ and $\gamma = \lambda_1 {v}_0$.
These parameters depend on the scalar quantities ${v}^2$ and $\rho({\bm r})$ whose fluctuations are small in the broken symmetry 
state. So, hereafter we consider these parameters as constants. 

It is now instructive to Fourier transform the set of Eqs.~(\ref{eqmdelrho})-(\ref{eqmdelvx}) in space and time.
Given a function $u({\bm r}, t)$, its Fourier transform in space and time is defined as 
\begin{equation}
 u({\bm q}, \omega) = \int_{-\infty}^{\infty} dt d{\bm r} e^{i \omega t} e^{-i{\bm q}\cdot{\bm r}} u({\bm r}, t).
 \label{fourdef}
\end{equation}
Using the above definition, we write the equations of motion for the fluctuations in the Fourier space as follow 
\begin{eqnarray}
 \left[ i\left( \omega - {v}_0 q_y \right) - \Gamma_{\rho}({\bm q}) \right] \delta\rho - i\bar\rho q_x \delta{v}_x 
   &=& -i \frac{\rho_o}{2\alpha_1} q_y \zeta_y, \notag \\ \label{eqfour1}\\
 i\sigma_1 q_x \delta\rho + \left[ -i\left( \omega - \gamma q_y \right) + \Gamma_L({\bm q}) \right] \delta{v}_x &=& 
     \frac{\rho_o}{\bar\rho} \zeta_x + f_x, \notag \\ \label{eqfour2}
\end{eqnarray}
where wave number dependent dampings are
\begin{eqnarray}
 \Gamma_{\rho}({\bm q}) &=& D_{\rho} q_x^2 + D_{\rho y} q_y^2, \\
 \Gamma_L({\bm q}) &=& D_L q_x^2 + D_y q_y^2 . \label{longdamp}
\end{eqnarray}
The normal modes of the pair of coupled Eqs.~(\ref{eqfour1})-(\ref{eqfour2}) are two propagating sound waves with complex 
eigenfrequencies
\begin{equation}
 \omega_{\pm} = c_{\pm}(\theta)q - i\Gamma_L \left[ \frac{{v}_{\pm}(\theta)}{2c_2(\theta)} \right] 
                                 - i\Gamma_{\rho} \left[ \frac{{v}_{\mp}(\theta)}{2c_2(\theta)} \right],
 \label{eigenfreq2}
\end{equation}
where $\theta$ is the angle between ${\bm q}$ and the direction of flock, i.e., $\hat y$-direction, and
\begin{eqnarray}
 c_{\pm}(\theta) &=& \frac{\gamma + {v}_0}{2} \cos\theta \pm c_2(\theta), \label{eqcpm} \\
 {v}_{\pm}(\theta) &=& c_2(\theta) \pm \frac{\gamma - {v}_0}{2} \cos\theta, \label{eqvpm} \\
 c_2(\theta) &=& \sqrt{\frac{1}{4} (\gamma - {v}_0)^2 \cos^2\theta + \sigma_1 \bar\rho \sin^2\theta}. \label{eqc2}
\end{eqnarray}
Solving the linear set of Eqs.~(\ref{eqfour1})-(\ref{eqfour2}) for $\delta\rho({\bm q},\omega)$ and $\delta{v}_x({\bm q},\omega)$, 
we obtain
\begin{eqnarray}
 \left[
 \begin{array}{cc}
 \delta\rho({\bm q},\omega) \\
 \delta{v}_x({\bm q},\omega)
 \end{array}
 \right] = \left[
 \begin{array}{cc}
 G_{\rho\rho}({\bm q},\omega) & G_{\rho L}({\bm q},\omega) \\
 G_{L \rho}({\bm q},\omega)   & G_{LL}({\bm q},\omega)
 \end{array}
 \right] \left[
 \begin{array}{cc}
 i\rho_o \zeta_y q_y / 2\alpha \\
 f_x + \rho_o \zeta_x / \bar\rho
 \end{array}
 \right] \notag \\ \label{delrhovx}
\end{eqnarray}
where the propagators are 
\begin{widetext}
\begin{eqnarray}
 G_{\rho\rho}({\bm q},\omega) &=& \frac{-i\left(\omega - \gamma q \cos\theta\right) + \Gamma_L({\bm q})}
   {\left[\omega-c_+(\theta)q\right]\left[\omega-c_-(\theta)q\right] + i\omega\left[\Gamma_{\rho}({\bm q})+\Gamma_L({\bm q})\right]
    - iq \cos\theta\left[\gamma\Gamma_{\rho}({\bm q}) + {v}_0\Gamma_L({\bm q})\right]}, 
 \label{Grhorho} \\
 G_{\rho L}({\bm q},\omega) &=& \frac{i \bar\rho q \sin\theta}
   {\left[\omega-c_+(\theta)q\right]\left[\omega-c_-(\theta)q\right] + i\omega\left[\Gamma_{\rho}({\bm q})+\Gamma_L({\bm q})\right]
    - iq \cos\theta\left[\gamma\Gamma_{\rho}({\bm q}) + {v}_0\Gamma_L({\bm q})\right]}, 
 \label{GrhoL} \\
 G_{L \rho}({\bm q},\omega) &=& \frac{-i \sigma_1 q \sin\theta}
   {\left[\omega-c_+(\theta)q\right]\left[\omega-c_-(\theta)q\right] + i\omega\left[\Gamma_{\rho}({\bm q})+\Gamma_L({\bm q})\right]
    - iq \cos\theta\left[\gamma\Gamma_{\rho}({\bm q}) + {v}_0\Gamma_L({\bm q})\right]}, 
 \label{GLrho} \\
 G_{LL}({\bm q},\omega) &=& \frac{i\left(\omega - {v}_0 q \cos\theta\right) - \Gamma_{\rho}({\bm q})}
   {\left[\omega-c_+(\theta)q\right]\left[\omega-c_-(\theta)q\right] + i\omega\left[\Gamma_{\rho}({\bm q})+\Gamma_L({\bm q})\right]
    - iq \cos\theta\left[\gamma\Gamma_{\rho}({\bm q}) + {v}_0\Gamma_L({\bm q})\right]}. 
 \label{GLL} 
\end{eqnarray}

Using the expression given in Eqs.~(\ref{delrhovx})-(\ref{GLL}) and the correlations given in the main text,
we calculate correlation functions for the density and the velocity fields. Retaining upto lowest-order terms in $q$, 
we obtain density-density correlation function  
\begin{eqnarray}
 C_{\rho\rho}({\bm q},\omega) &=& \frac{\left(\omega-\gamma q\cos\theta\right)^2 \left[-\frac{\rho_o^2}{4\alpha_1^2}
    \zeta^2 q^2\cos^2\theta \delta(\omega)\right] + \bar\rho^2 q^2\sin^2\theta \left[\Delta + 
    \frac{\rho_o^2}{\bar\rho^2} \zeta^2 \delta(\omega)\right]}
   {\left[\omega-c_+(\theta)q\right]^2\left[\omega-c_-(\theta)q\right]^2 + \left\{ \omega\left[\Gamma_{\rho}({\bm q})+\Gamma_L({\bm q})\right]
    - q_y\left[\gamma\Gamma_{\rho}({\bm q}) + {v}_0\Gamma_L({\bm q})\right] \right\}^2 }, 
 \label{Crhorho1} 
\end{eqnarray}
and velocity-velocity correlation function
\begin{eqnarray}
 C_{{vv}}({\bm q},\omega) &=& \frac{-\sigma_1^2\zeta^2q^4\sin^22\theta \frac{\rho_o^2}{16\alpha_1^2} \delta({\omega}) +
  \left(\omega-{v}_0q\cos\theta\right)^2 \left[ \Delta + \frac{\rho_o^2}{\bar\rho^2} \zeta^2 \delta(\omega) \right] } 
   {\left[\omega-c_+(\theta)q\right]^2\left[\omega-c_-(\theta)q\right]^2 + \left\{ \omega\left[\Gamma_{\rho}({\bm q})+\Gamma_L({\bm q})\right]
    - q_y\left[\gamma\Gamma_{\rho}({\bm q}) + {v}_0\Gamma_L({\bm q})\right] \right\}^2 }. 
 \label{Cvv1} 
\end{eqnarray}
\end{widetext}
Given these Fourier transformed correlation functions, we proceed further to obtain the spatially Fourier transformed equal-time correlation 
functions for the density and the velocity fields. Neglecting the higher order fluctuations, we obtain the 
expressions for $C_{\rho\rho, vv}({\bm q},t)$, as also given in the main text :
\begin{eqnarray}
 C_{\rho\rho}({\bm q},t) = 
    \frac{1}{q^2} \left\{ \frac{\zeta^2 \rho_o^2 a_{\rho}(\theta)}{b(\theta)q^2+d(\theta)} + \Delta A_{\rho}(\theta) \right\},
 \label{appCrhorho2} \\
 C_{{vv}}({\bm q},t) = 
     \frac{1}{q^2} \left\{ \frac{\zeta^2 \rho_o^2 a_{v}(\theta)}{b(\theta)q^2+d(\theta)} + \Delta A_{v}(\theta) \right\},
 \label{appCvv2}
\end{eqnarray}
where
\begin{eqnarray}
 b(\theta) &=& \cos^2\theta \{ \gamma\left(D_{\rho}\sin^2\theta + D_{\rho y}\cos^2\theta\right) \notag \\
      && \qquad \qquad + {v}_0 \left(D_L\sin^2\theta + D_y\cos^2\theta\right) \}^2, \label{defb} \\
 d(\theta) &=& \gamma{v}_0\cos^2\theta - \sigma_1\bar\rho\sin^2\theta, \label{defd} \\ 
 p_{\pm}(\theta) &=& 2c_2 \{ c_{\pm} [(D_L+D_{\rho})\sin^2\theta + (D_y+D_{\rho y})\cos^2\theta] \notag \\ 
  && \qquad - \cos\theta [({v}_0D_L+\gamma D_{\rho})\sin^2\theta \notag \\ 
  && \qquad \qquad \qquad + ({v}_0D_y + \gamma D_{\rho y}) \cos^2\theta] \}, \\
 s_{\pm}(\theta) &=& (c_{\pm} - {v}_0\cos\theta)^2 / 2\pi, \\
 s(\theta) &=& \bar\rho^2\sin^2\theta / 2\pi, 
\end{eqnarray}
\begin{eqnarray}
 a_{\rho}(\theta) &=& \sin^2\theta / 2\pi, \\
 A_{\rho}(\theta) &=& s(\theta)\left[ \frac{1}{p_+(\theta)} + \frac{1}{p_-(\theta)} \right], \\
 a_{v}(\theta) &=& {v}_0^2\cos^2\theta / 2\pi\bar\rho^2, \\
 A_{v}(\theta) &=& \left[ \frac{s_+(\theta)}{p_+(\theta)} + \frac{s_-(\theta)}{p_-(\theta)} \right]. 
\end{eqnarray}

%%%%%%%%%%%%%%%%%%%%%%%%%%%%%%%%%%%%%%%%%%%%%%%%%%%%%%%%%%%%%%%%%%%%%%%%%%%%%%%%%%%%%%%%%%%%%%%%%%%%%%%%%%%%%%%%%%%%%%%%%%%%%%%%%%%%%%%%%%%%%%%%%%

\section{Effect of nonlinear terms} \label{appnonlinear}
Linearised hydrodynamics suggests that $C_{\rho\rho, vv} \sim 1/q^4$ for small $q$ if $d(\theta) = 0$, otherwise $C_{\rho\rho, vv} 
\sim 1/q^2$. It is clear from Eq.~(\ref{defd}) that $d(\theta)$ cannot vanish for  $u = \gamma v_0 / \sigma_1 \bar\rho < 0$. Also 
for the case $u>0$, $d(\theta)$ vanishes only for $\theta = \theta_c$ and $\pi - \theta_c$, where $\theta_c = \tan^{-1}\sqrt{u}$. 
Therefore, the appearance of the $1/q^4$ divergence, as predicted by the linearized calculation, is relevant only for the case 
$u>0$, and that too along the two specific directions mentioned above. 

Let us first discuss about the physical significance of the sign of $u$. The sign of $u=\gamma v_0/\sigma_1\bar\rho$ depends on the 
microscopic parameters of the concerned model. In general for Vicsek-like models, $u>0$ which infers that the pure modes for the 
velocity and the density fields are more or less {\it in phase}. However, there can be situation where these pure modes are almost 
in {\it opposite phases} inferring $u < 0$, which we shall not discuss here. Since, our numerical model in this paper is a modification 
of the Vicsek model (VM), we stress that $u>0$ in our model. Then the immediate question arises if the higher order divergence predicted 
by the linearized calculation destabilises the QLRO that we claim for. Here we ascertain that the neglected nonlinear terms of the 
hydrodynamic EOMs make the existence of the QLRO possible in the system. 

A common method  to incorporate the effect of different nonlinearities is to first find out the dominating terms by 
`power counting' \cite{tonertu1995, tonertu1998, toner2012, toner2018}. Unfortunately, that technique is not useful in our case, as it 
predicts all the nonlinearities as equally significant in 2D. Furthermore, a complete renormalisation group analysis is {\it not 
practically feasible} \cite{toner2018}; However, if we assume that $\lambda_1$ term of the velocity EOM in the main text is the most 
relevant nonlinearity, then one can calculate the exact exponents for the model. Though this seems like an {\it ad hoc} assumption, a similar 
assumption practically works well in the presence of ${\bm f}$ (annealed) noise {\it only}. Therefore, one can proceed with this assumption for 
the quenched case also, as recently done by Toner {\it et al.} \cite{toner2018} but considering few more extra terms in the EOMs \cite{toner2012}. 
Though these extra terms are allowed by the symmetry of the system, but these terms do not change the system phenomenology further \cite{toner2012}. 
Therefore, the calculation done in Ref.~\cite{toner2018} equally holds for our model also, which predicts that 
the $\lambda_1$ nonlinearity renormalizes the diffusivities as $1/q$ for small $q$. Therefore, from Eq.~(\ref{defb}) we can see 
that $b(\theta) \sim 1/q^2$. Hence, the product $b(\theta) q^2 $ approaches a finite value.
Therefore, though the linearized calculation predicts highly anisotropic $C_{\rho\rho, vv}$, the convective nonlinearity suppresses 
that anisotropy, and makes the QLRO possible in our system. 

The effect of nonlinearities in suppression of the higher order fluctuations predicted by the linearized theory can 
also be understood by the number fluctuation $\Delta N = \sqrt{\langle N^2 \rangle - \langle N \rangle ^2}$ in the AMQR. 
We note that in the clean system, $\Delta N \sim \langle N \rangle ^{\kappa}$ with $\kappa > 1/2$. In the presence of the 
RQRs also, the system shows giant number fluctuation, but $\kappa$ varies with $c_r$. We note that the number fluctuation 
for small $c_r$ is higher than the clean case. However, the fluctuation decreases with further increase in $c_r$, which is 
not allowed by Eq.~(\ref{appCrhorho2}) as the quenched terms should dominate over the annealed terms with increasing density 
of inhomogeneity. This observation suggests that the neglected nonlinearities play pivotal role in 
stabilizing ordered state in the system. Moreover, the scaling of the order parameter ${\rm V}$ with system size, as described 
in the main text, ensures us about the existence of the QLRO (and not the LRO) in the presence of quenched inhomogeneities.

%%%%%%%%%%%%%%%%%%%%%%%%%%%%%%%%%%%%%%%%%%%%%%%%%%%%%%%%%%%%%%%%%%%%%%%%%%%%%%%%%%%%%%%%%%%%%%%%%%%%%%%%%%%%%%%%%%%%%%%%%%%%%%%%%%%%%%%%%%%%%%%%%%

\begin{figure}[b]
  \includegraphics[width=0.9\linewidth]{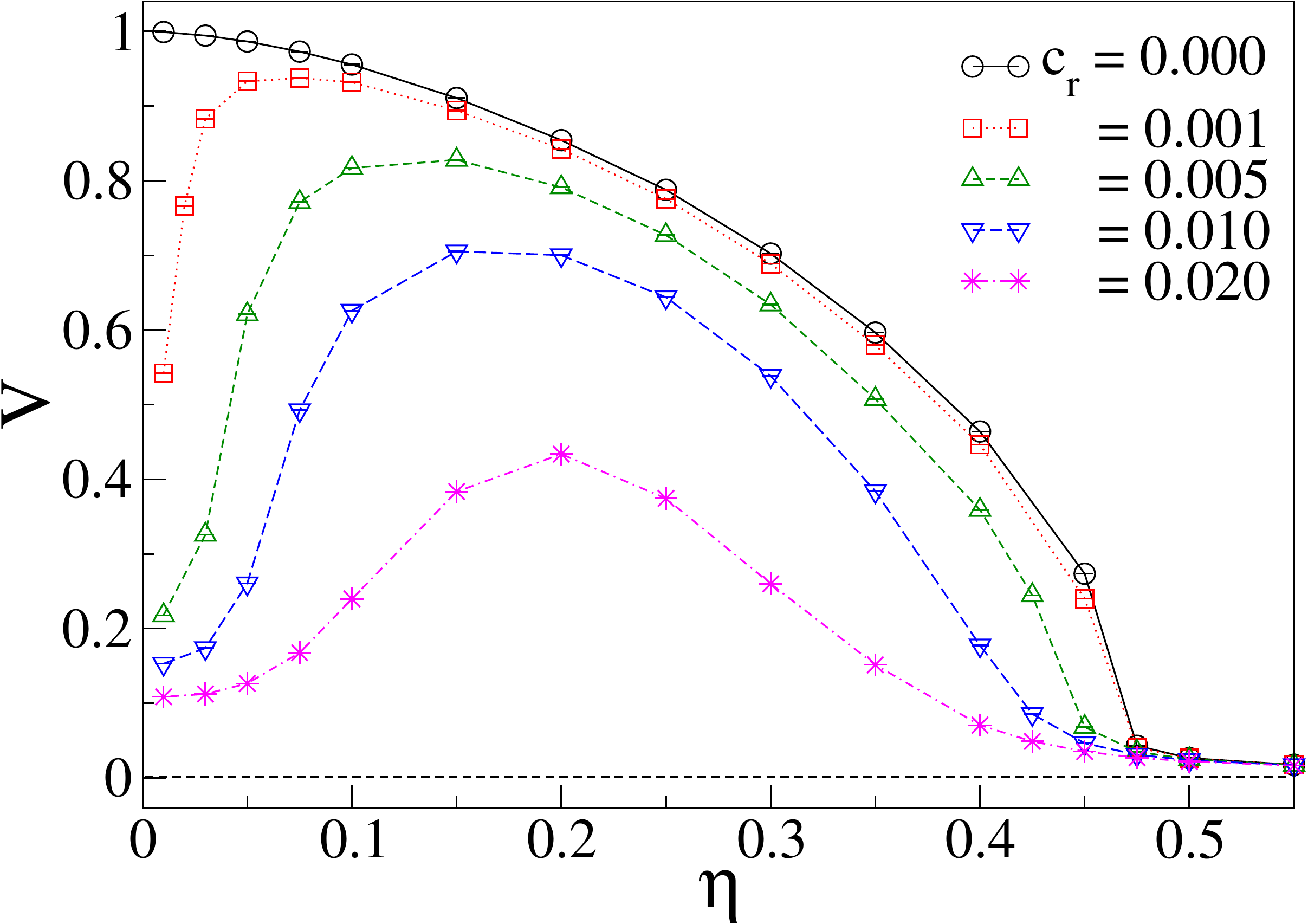}
  \caption{  Plot of average normalized velocity versus noise strength shown for $L=100$ and different density
            of the rotators. While ${\rm V}$ decreases monotonically with increasing $\eta$ for the clean system, in the presence
            of the rotators, a certain amount of noise maximizes the ordering. }
\label{figoptnoise}
\end{figure}

%%%%%%%%%%%%%%%%%%%%%%%%%%%%%%%%%%%%%%%%%%%%%%%%%%%%%%%%%%%%%%%%%%%%%%%%%%%%%%%%%%%%%%%%%%%%%%%%%%%%%%%%%%%%%%%%%%%%%%%%%%%%%%%%%%%%%%%%%%%%%%%%%%

\section{Optimal noise} \label{appoptnoise}
In Fig.~\ref{figoptnoise}, we show the variation of ${\rm V}$ with $\eta$ in the AMQR. 
In the clean system, ${\rm V}$ decays monotonically with increasing $\eta$. Surprisingly, in the presence of the RQRs, a
certain amount of noise facilitates flocking, and below that the ordering reduces again. This happens because, in the presence of the
RQRs, the system must have a non-zero noise to transfer the information of one sub-flock to another. Similar phenomenon has earlier
been reported in Ref.~\cite{chepizhko2013}.

%%%%%%%%%%%%%%%%%%%%%%%%%%%%%%%%%%%%%%%%%%%%%%%%%%%%%%%%%%%%%%%%%%%%%%%%%%%%%%%%%%%%%%%%%%%%%%%%%%%%%%%%%%%%%%%%%%%%%%%%%%%%%%%%%%%%%%%%%%%%%%%%%%

%%%%%%%%%%%%%%%%%%%%%%%%%%%%%%%%%%%%%%%%%%%%%%%%%%%%%%%%%%%%%%%%%%%%%%%%%%%%%%%%%%%%%%%%%%%%%%%%%%%%%%%%%%%%%%%%%%%%%%%%%%%%%%%%%


\begin{thebibliography}{10}
%
\bibitem{vicsek1995} T. Vicsek, A. Czir\'ok, E. Ben-Jacob, I. Cohen, and O. Shochet, Phys. Rev. Lett. {\bf 75}, 1226 (1995).
%
\bibitem{revmarchetti2013} M. C. Marchetti, J. F. Joanny, S. Ramaswamy, T. B. Liverpool, J. Prost, Madan Rao, and R. Aditi Simha, 
Rev. Mod. Phys. {\bf 85}, 1143 (2013).
%
\bibitem{revtoner2005} J. Toner, Y. Tu, and S. Ramaswamy, Ann. Phys. {\bf 318,} 170 (2005).
%
\bibitem{revramaswamy2010} S. Ramaswamy, Annu. Rev. Cond. Matt. Phys. {\bf 1}, 323 (2010).
% 
\bibitem{revvicsek2012} T. Vicsek and A. Zafeiris, Phys. Rep. {\bf 517}, 71 (2012).
%
\bibitem{revcates2012} M. E. Cates, Rep. Prog. Phys. {\bf 75}, 042601 (2012).
%
\bibitem{tonertu1995} J. Toner and Y. Tu, Phys. Rev. Lett. {\bf 75}, 4326 (1995).
%
\bibitem{tonertu1998} J. Toner and Y. Tu, Phys. Rev. E {\bf 58}, 4828 (1998).
%
\bibitem{shradha2010} S. Mishra, A. Baskaran and C. Marchetti, Phys. Rev. E {\bf 81}, 061916 (2010).
%
\bibitem{chate2008} H. Chat{\'e}, F. Ginelli, G. Gr{\'e}goire and F. Raynaud, Phys. Rev. E {\bf 77}, 046113 (2008).
%
\bibitem{morin2017} A. Morin, N. Desreumaux, J. B. Caussin, and D. Bartolo, Nat. Phys. {\bf 13}, 63 (2017).
%
\bibitem{chepizhko2013} O. Chepizhko, E. G. Altmann and F. Peruani, Phys. Rev. Lett. {\bf 110}, 238101 (2013).
%
\bibitem{marchetti2017} D. Yllanes, M. Leoni and M. C. Marchetti, New J. Phys. {\bf 19}, 103026 (2017).
%
\bibitem{quint2015} D. A. Quint and A. Gopinathan, Phys. Biol. {\bf 12}, 046008 (2015).
%
\bibitem{sandor2017} Cs. S{\'a}ndor, A. Lib{\'a}l, C. Reichhardt and C. J. O. Reichhardt, Phys. Rev. E {\bf 95}, 032606 (2017).
%
\bibitem{reichhardt2017} C. J. O. Reichhardt and C. Reichhardt, Nat. Phys. {\bf 13}, 10 (2017).
%
\bibitem{imry1975} Y. Imry and S. K. Ma, Phys. Rev. Lett. {\bf 35}, 1399 (1975).
%
\bibitem{grinstein1976} G. Grinstein and A. Luther, Phys. Rev. B {\bf 13}, 1329 (1976).
%
\bibitem{chaikin} P. M. Chaikin and T. C. Lubensky, {\it Principles of Condensed Matter Physics} (Cambridge University Press, Cambridge, 1998).
%
\bibitem{voigt} J. J. Olivero and R. L. Longbothum, J. Quant. Spectrosc. Radiat. Transfer {\bf 17}, 233 (1977).
%
\bibitem{aparna2008} A. Baskaran, J. W. Dufty and J. J. Brey, Phys. Rev. E {\bf 77}, 031311 (2008).
%
\bibitem{bertin2006} E. Bertin, M. Droz and G. Gr{\'e}goire, Phys. Rev. E {\bf 74}, 022101 (2006).
%
\bibitem{bertin2009} E. Bertin, M. Droz and G. Gr{\'e}goire, J. Phys. A: Math. Theor. {\bf 42}, 445001 (2009).
%
\bibitem{ihle2011} T. Ihle, Phys. Rev. E {\bf 83}, 030901(R) (2011).
%
\bibitem{toner2018} J. Toner, N. Guttenberg and Y. Tu, arXiv:1805.10324v1 [cond-mat.stat-mech]; arXiv:1805.10326v1 [cond-mat.stat-mech].
%
\bibitem{couzin2008} D. J. T. Sumpter, J. Krause, R. James, I. D. Couzin and A. J. W. Ward, Curr. Biol. {\bf 18}, 1773 (2008).
%
\bibitem{jolles2017} J. W. Jolles, N. J. Boogert, V. H. Sridhar, I. D. Couzin and A. Manica, Curr. Biol. {\bf 27}, 2862 (2017).
%
\bibitem{krause2018} L. Snijders, R. H. J. M. Kurvers, I. W. Ramnarine and J. Krause, Nat. Ecol. Evol. {\bf 2}, 1610 (2018).
%
\bibitem{puckett2018} J. G. Puckett, A. R. Pokhrel and J. A. Giannini, Sci. Rep. {\bf 27}, 7587 (2018).
%
\bibitem{toner2012} J. Toner, Phys. Rev. E {\bf 86}, 031918 (2012).
%
\end{thebibliography}
\end{document}